\documentclass{article}
\usepackage{spconfa4}
\newcommand{\norm}[1]{\left\lVert#1\right\rVert}

\def\where{\text{ where }}
\def\ba{{\mathbf{a}}}     
  \def\bg{{\mathbf{g}}} \def\bh{{\mathbf{h}}}
     
     \def\bp{{\mathbf{p}}}
  \def\br{{\mathbf{r}}}

\def\bA{{\mathbf{A}}}

\def\T{^\mathsf{T}}

\newcommand*\xbar[1]{%
  \hbox{%
    \kern 0.1em
    \vbox{%
      \hrule height 0.5pt % The actual bar
      \kern0.3ex%         % Distance between bar and symbol
      \hbox{%
        \kern-0.0em%      % Shortening on the left side
        \ensuremath{#1}%
        \kern-0.0em%      % Shortening on the right side
      }%
    }%
    \kern 0.0em
  }% hbox
} 

\usepackage{cite}
\usepackage{amsmath,amssymb,amsfonts}
\usepackage{graphicx}
\usepackage{multirow}
\usepackage{slashbox}
\title{RGI-Net: 3D Room Geometry Inference from Room Impulse Responses With Hidden First-Order Reflections}
\name{Inmo Yeon and Jung-Woo Choi\thanks{This work was supported by the National Research Foundation of Korea (NRF) grant funded by the Ministry of Science and ICT of Korea government (MSIT) (No. RS-2024-00337945), the BK21 FOUR program through the NRF grant funded by the Ministry of Education of Korea government (MOE), and Mobile eXperience (MX) Business, Samsung Electronics Co., Ltd.}}
\address{School of Electrical Engineering, KAIST, Daejeon, South Korea
}
\begin{document}
%\ninept
%
\maketitle
\begin{abstract}
Room geometry is important prior information for implementing realistic 3D audio rendering. For this reason, various room geometry inference (RGI) methods have been developed by utilizing the time-of-arrival (TOA) or time-difference-of-arrival (TDOA) information in room impulse responses (RIRs). However, the conventional RGI technique poses several assumptions, such as convex room shapes, the number of walls known in priori, and the visibility of first-order reflections. In this work, we introduce the RGI-Net which can estimate room geometries without the aforementioned assumptions. RGI-Net learns and exploits complex relationships between low-order and high-order reflections in RIRs and, thus, can estimate room shapes even when the shape is non-convex or first-order reflections are missing in the RIRs. RGI-Net includes the evaluation network that separately evaluates the presence probability of walls, so the geometry inference is possible without prior knowledge of the number of walls.
\end{abstract}
\section{Introduction}
\label{sec:intro}

Sound propagation in a room is affected by various physical phenomena involved with room geometries, such as specular and diffuse reflection, diffraction, and scattering. Therefore, knowing the room geometry can improve the performance of many acoustic problems including source localization \cite{intro_SourceLocalization}, dereverberation \cite{intro_Dereverberation}, and source separation \cite{intro_RakingTheCocktailParty}. 

Sound propagation altered by room geometry is captured in room impulse responses (RIRs) in the form of various features. One of the representative features is the time-of-arrival (TOA) identifiable from the propagation delay of its distinct peaks. Therefore, a lot of room geometry inference (RGI) studies have utilized TOA information \cite{6_antonacci, 10_dokmanic, 5_nastasia, 15_remaggi, 25_baba, 28_lovedee}.
In particular, Antonacci \textit{et al.} \cite{6_antonacci} localized 2D reflectors by converting a set of TOAs into ellipses. Two focal points of an ellipse correspond to a sound source and microphone, and the common tangent line across multiple ellipses can represent the room boundary. 
Baba \textit{et al.} \cite{25_baba} extended ellipse-based method using stack-line detection in stacked RIRs. Common reflection lines were translated into real- and image-microphone positions, which determine room boundary. 
Lovedee-Turner and Murphy \cite{28_lovedee} demonstrated that the RGI of complex rooms, e.g., non-convex rooms, is possible based on RIRs acquired at multiple positions. They estimated all candidate walls and then filtered the estimated walls using a three-step validation process: path validation, line-of-sight (LOS) boundary validation, and closed geometry validation. The remaining walls constitute the final room geometry. However, the microphone array must be placed in a restricted area to secure the LOS condition for all the walls, and the acquired RIRs must include first-order reflections from the walls. Despite their success in estimating complex-shaped rooms, positioning sound sources at multiple locations makes it less practical.

%% Model
\begin{figure*}[t]
    \centering
    \includegraphics[width=\textwidth]{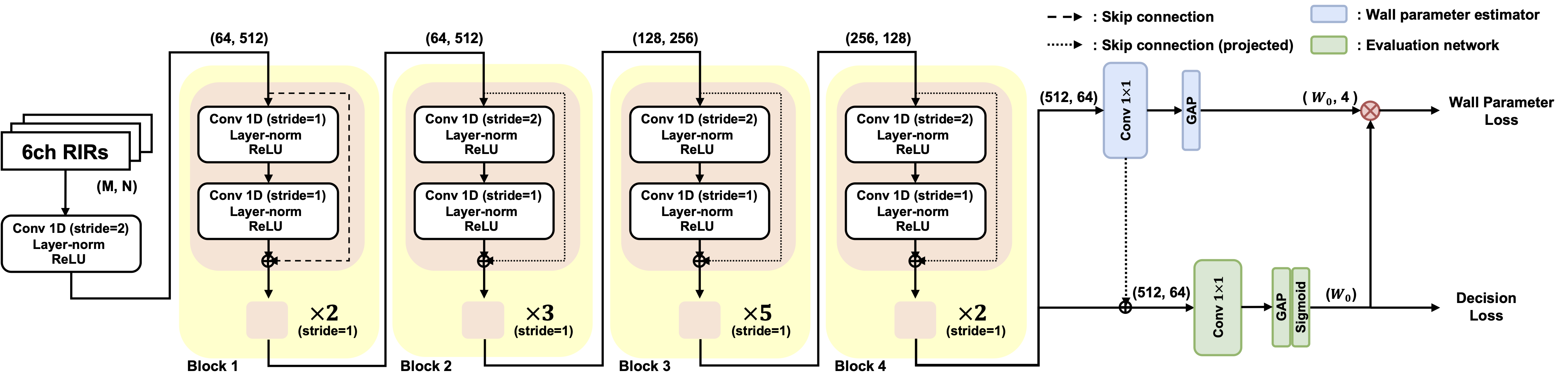}
    \caption{Overview of the RGI-Net architecture. $M$, $N$, and $W_0$ denote the number of channels and temporal length of RIRs, and the maximum number of walls, respectively.}
    \label{fig:model}
    \vspace{-0.4cm}
\end{figure*}

Most previous studies assume that first-order reflections from every wall are visible. For rooms with complex shapes, however, first-order reflection visibility cannot be ensured in one measurement with a compact microphone array. The fundamental remedy might be to utilize high-order reflections and analyze the complex relationship between low- and high-order reflections. 
Therefore, recent studies have demonstrated the potential of deep neural networks (DNNs) for high-level feature extraction from RIR \cite{17_yu, 18_poschadel, tuna_datadriven, yeon_ica, neural_acoustic_fields, floorplan_reconstruction}. Yu and Kleijn \cite{17_yu}, Poschadel \textit{et al.} \cite{18_poschadel}, and Tuna \textit{et al.}\cite{tuna_datadriven} designed a DNN to extract the complex relationship between the temporal peaks of an RIR and estimated the geometrical parameters of a room. 
However, these DNN-based RGI studies have a limitation in that they cannot handle changes in the number of walls caused by various room types, as they have focused only on shoebox rooms.
In our previous study \cite{yeon_ica}, we attempted to address this limitation. However, the study only considered the LOS condition where all walls are visible from the audio device. Moreover, a spherical microphone array with 32 microphone capsules was used as the audio device, which makes the approach less practical. 

In this study, we propose a DNN-based RGI model, RGI-Net, that can infer geometries of various rooms. The proposed model has two major contributions compared to conventional RGI techniques.
First, RGI-Net is designed to infer room geometry without prior knowledge of the number of walls. 
Second, RGI-Net is capable of inferring the room geometries, even when they do not satisfy the LOS conditions. The exploitation of high-order reflections is demonstrated by the visualization of temporal activation maps.

\section{Problem Statement}
The RGI problem can be defined as identifying $W$ walls that compose a room based on the measured RIRs. In 3D space, each wall is described as a plane, which can be expressed as a set ($\mathcal{A}_w$) of points ($\br=[ x, ~y,~ z,~ 1 ]\T$) in homogeneous coordinates, satisfying the equation of a plane.
\begin{equation}\label{plane_eq}
    \mathcal{A}_w = \{\br \in \mathbb{R}^4\; \vert \; \br\T \ba_w = 0\},
\end{equation}
where the vector $\ba_w = [a_{w1},~ a_{w2},~ a_{w3},~ a_{w4}]\T$ includes wall parameters characterizing the $w$th wall ($w=\{1,\cdots,W\}$).
The objective of RGI can be accomplished by determining $\ba_w$ constituting a room.
Hereafter, we describe the architecture of RGI-Net to estimate wall parameters without prior information on the number of walls ($W$).

\section{Proposed Method}

%% Table 1
\begin{table*}[t]
\caption{Performance of RGI-Net on different room geometries at low- and high-noise levels.}
\centering
\resizebox{\textwidth}{!}
{
\footnotesize
\begin{tabular}{l|cc|cc|cc|cc|cc|cc}
\hline
\multicolumn{1}{c|}{Room type} &
  \multicolumn{2}{c|}{Total} &
  \multicolumn{2}{c|}{Shoebox} &
  \multicolumn{2}{c|}{Pentagonal} &
  \multicolumn{2}{c|}{Hexagonal} &
  \multicolumn{2}{c|}{L-LOS} &
  \multicolumn{2}{c}{L-NLOS} \\ \hline
Number of RIRs &
  \multicolumn{2}{c|}{500} &
  \multicolumn{2}{c|}{100} &
  \multicolumn{2}{c|}{100} &
  \multicolumn{2}{c|}{100} &
  \multicolumn{2}{c|}{100} &
  \multicolumn{2}{c}{100} \\ \hline
Background noise level   & Low  & High & Low  & High & Low  & High & Low  & High & Low  & High & Low  & High \\ \hline \hline
\begin{tabular}[c]{@{}l@{}}$ACC_W$ (\%)\\ (ROC--AUC, \%)\end{tabular} &
  \begin{tabular}[c]{@{}c@{}}99.95\\ (99.99)\end{tabular} &
  \begin{tabular}[c]{@{}c@{}}99.72\\ (99.98)\end{tabular} &
  100 &
  100 &
  99.88 &
  99.38 &
  99.88 &
  99.75 &
  100 &
  100 &
  100 &
  99.50 \\ \hline
$\Delta d$ (m)           & 0.10 & 0.16 & 0.07 & 0.11 & 0.08 & 0.14 & 0.08 & 0.14 & 0.11 & 0.15 & 0.16 & 0.28 \\ \hline
$\Delta \theta$ (degrees) & 1.89 & 3.33 & 1.68 & 1.95 & 2.10 & 3.69 & 2.46 & 3.62 & 1.39 & 2.61 & 1.83 & 4.76 \\ \hline
\end{tabular}%
}
\label{tab:basic_results}
\vspace{-0.4cm}
\end{table*}

\subsection{RGI-Net Architecture}
The proposed network comprises three sub-networks: a feature extractor, a wall parameter estimator, and an evaluation network. The feature extractor extracts appropriate features related to room geometries from multichannel RIRs. The $M$-channel RIRs $\bg \in \mathbb{R}^{M \times N }$ of $N=1024$ in length are processed by the convolutional layer of kernel size 9 and stride 2 and fed into the feature extractor.
As shown in Fig. \ref{fig:model}, the ResNet \cite{resnet} used as a feature extractor consists of four main blocks. As the signal passes through 1D convolution layers, the number of channels increases while the length of the feature map decreases. The convolution layers extract interchannel and temporal features through the multichannel kernel of size 5 and stride 1, except for the first layer of blocks 2, 3, and 4, where stride is 2. Each convolutional layer is followed by layer-norm \cite{layernorm} and rectified linear unit (ReLU) activation.

The wall parameter estimator is the combination of the 1$\times$1 convolution and global average pooling (GAP) layers, which mixes and summarizes the extracted features, respectively, to obtain a set of wall parameter estimations ($\hat{\ba}_w$). Irrespective of the number of existing walls ($W$), the estimator is designed to generate $W_0$ wall parameter candidates $\hat \bA_c =[\hat \ba_1,~ \cdots, ~\hat \ba_{ W_0}]\T \in \mathbb{R}^{ W_0 \times 4}$. A well-trained wall parameter estimator would generate near-zero vectors ($\norm{\hat \ba_w}_2\approx 0$) for nonexistent walls. However, to promote the detection of fake wall parameters, we additionally incorporate the evaluation network that evaluates and outputs the confidence of estimated parameters. In this sub-network, the presence probability $\hat{\bp} = [\hat{p}_1, ~\cdots,~ \hat{p}_{W_0}]\T \in \mathbb{R}^{W_0}$ of $W_0$ wall candidates is generated through the sigmoid function by considering both features of the feature extractor and wall parameter estimator.
During training, the final wall parameters are estimated by multiplying the output of the wall parameter estimator with the wall presence probability obtained from the evaluation network ($\hat{\bA} = \mathrm{Diag}(\hat \bp) \hat{\bA}_c $). However, during inference, the binary decision of true walls is made by hard-thresholding $\hat \bp$ with a threshold $0.9$, which is determined by considering the true-positive rate (TPR) and false-positive rate (FPR).

\subsection{Loss Function}
In all the experiments conducted, we set $W_0 =8 \ge W$, and ground truth (GT) wall parameters and presence probabilities for nonexistent walls were initialized with zeroes. As the loss function for measuring the similarity between GT and estimated wall, angular loss between two flattened parameter vectors $\hat{\bh} = \mathrm{flatten}(\hat{\bA})$ and ${\bh} = \mathrm{flatten}({\bA}) \in \mathbb{R}^{4W_0}$ was employed. That is, 
    \begin{equation}\label{angular_loss}
    L_{ang} = 1 - {\cos^2\theta},
    \where
    \cos\theta = \frac{\hat{\bh} \cdot \bh}{\|{\hat{\bh}}\|_2 \|{\bh}\|_2}.
    \end{equation}
Along with the angular loss, we used a decision loss $L_{d}$ involved with the wall presence probability. The decision loss is defined as the binary cross entropy (BCE) between the GT and estimated probabilities: $p_w$ and $\hat p_w$.
To cope with the order mismatch in the GT and estimated walls, the permutation invariant training technique \cite{pit} was employed during training.
For training, the Adam \cite{adam} optimizer and cosine annealing learning rate scheduler \cite{scheduler} were used with the maximum learning rate of $10^{-3}$.

\section{Experimental Results and Analysis}
\subsection{Dataset}
Although various measured RIRs have been released, their microphone-speaker configurations are not compatible with the compact audio device considered in this study. 
To train the model using RIRs obtainable from a compact audio system, we constructed an RIR dataset simulated using a circular microphone array with six omnidirectional microphones arranged on a circle of 0.05~m radius and a single loudspeaker centered in the array. The device was randomly positioned between $[1,~2]$~m from the floor and within 70\% space defined by equally scaling down from every vertex of a given floorplan of the room.
In the simulated rooms, the floor and ceiling were parallel to each other and perpendicular to the other walls. 
Four different room types were considered: shoebox, pentagonal, hexagonal, and L-type rooms. Unlike the other types, L-type rooms can be categorized into L-LOS and L-NLOS (non-line-of-sight) types, depending on whether the device can capture LOS from all walls or not.
The rooms were horizontally rotated within the range of $[0, ~360]^\circ$ to reflect possible angular rotations of the device.

The RIRs were simulated by `Pyroomacoustics' \cite{pyroomacoustics} %, which can model the RIRs of general polyhedral rooms. 
%For the train dataset, the RIRs were simulated 
using the image source method (ISM) at a sampling rate of 8\:kHz. The absorption coefficient of each room was randomly selected within $[0.1, ~0.3]$. The loudspeaker and microphones kept a consistent distance, so we trimmed out the direct part of the RIRs.
Two datasets with low- and high-noise levels were constructed by adding white Gaussian noise to RIRs. The noise level was adjusted to maintain a signal-to-noise ratio (SNR) within the ranges of $[20,~30]$ and $[10,~20]$~dB for low- and high-noise level datasets, respectively. In the high-noise level dataset, the noise level was sufficiently high to mask the peaks from third- or higher-order reflections. The resultant train dataset contains 600k RIRs, including 1k validation data.

%% FIG reconstruction
\begin{figure}[t]
    \centering
    \includegraphics[width=\columnwidth]{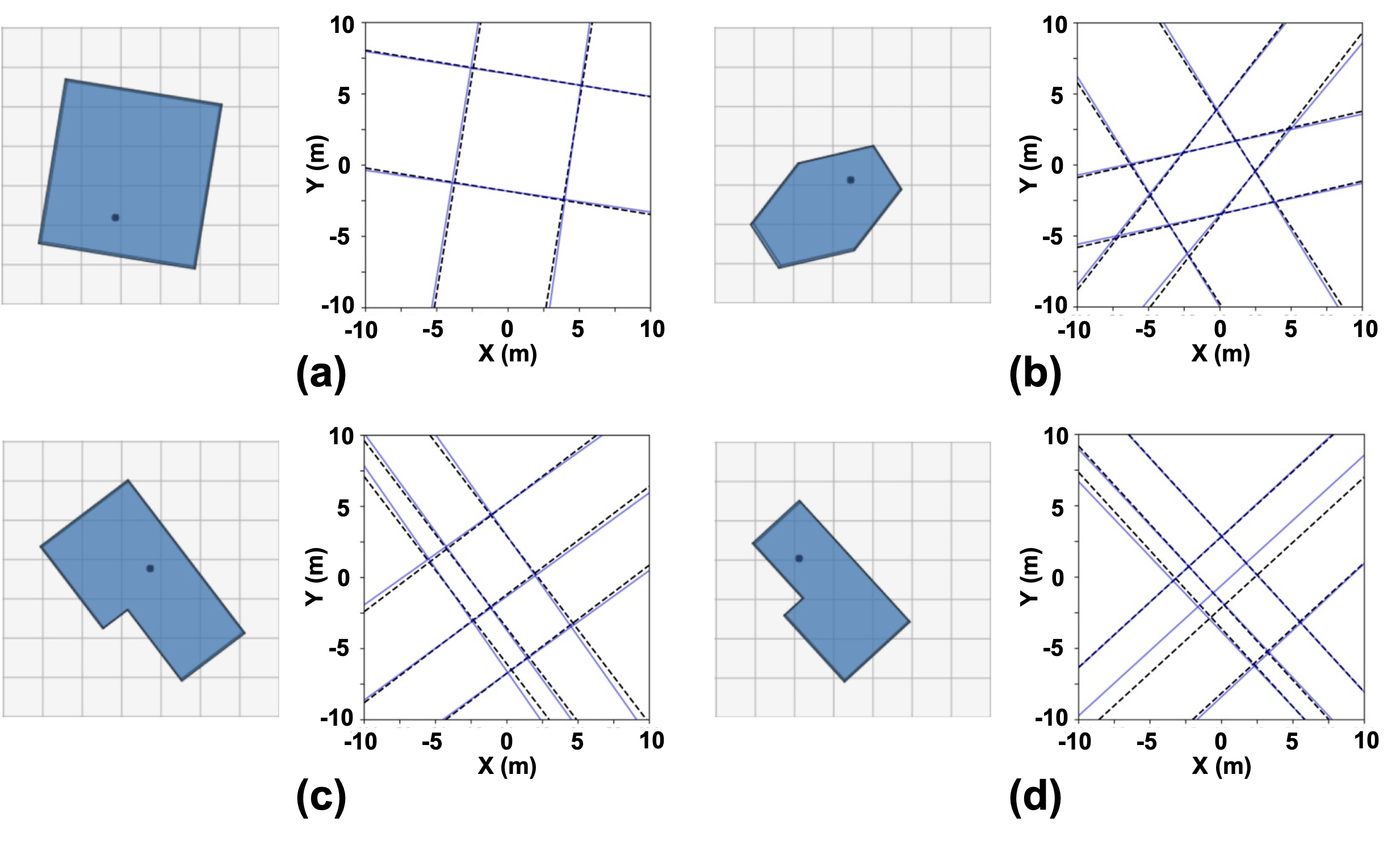}
    \caption{Top view of rooms reconstructed from estimated wall parameters. Since four distinct L-shaped rooms can be formed by the estimated planes, (c) and (d) were reconstructed considering the GT room shapes. The black dot (left) denotes the position of an audio device. The black dashed lines and blue solid lines (right) correspond to reconstructed walls from the GT and inferred wall parameters, respectively.}
    \label{fig:reconstruction}
    \vspace{-0.5cm}
\end{figure}

\subsection{Evaluation Metrics}
The performance of the proposed model was verified using three types of evaluation metrics: accuracy of wall presence estimation ($ACC_W$), distance error ($\Delta d$), and dihedral angle ($\Delta \theta$). The $ACC_W$ is defined as the number of correctly estimated walls normalized by the total number of walls. Since there is a hard-thresholding operation during inference, $ACC_W$ can vary according to the threshold value. To achieve the threshold-independent evaluation of the wall presence probability, we also employed the area under the curve of the receiver operating characteristic (ROC--AUC) \cite{roc_auc1}.
The $\Delta d$ indicates the difference between the shortest distances from the device to the GT and estimated wall \cite{5_nastasia,25_baba,28_lovedee, tuna_datadriven}, while the $\Delta \theta$ represents the angle between the normals of the GT and inferred walls \cite{25_baba,28_lovedee, tuna_datadriven}. These errors were calculated only for the walls that satisfy $\mathrm{AND}(\hat p_w, p_w)=1$ to avoid the error calculation for phantom walls.

\subsection{Experimental Results}
The RGI results on two different noise levels are listed in Table \ref{tab:basic_results}. The overall result shows that only the existence of 2 and 11 walls out of 4k candidate walls was misclassified in the low- and high-noise datasets, respectively. Therefore, the $ACC_W$ of the low- and high-noise datasets are 99.95\% and 99.72\%, respectively. The average errors are approximately 10~cm and 1.9$^\circ$ for the low-noise dataset. 
RGI-Net shows good performance for most room types, but the most significant error variation between low- and high-noise levels is observed in L-NLOS rooms. 
Fig. \ref{fig:reconstruction} depicts the comparison between GT and inferred rooms based on the high-noise dataset. The left part of each result shows the location of the audio device (black dot), which was set as the origin of coordinates during inference. These results show that most of the inferred walls (blue solid lines) are very close to the GT walls (black dashed lines), while the invisible wall shown in Fig. \ref{fig:reconstruction}(d) exhibits a more significant error than the other walls in the L-NLOS room. This is because the strong background noise masks high-order reflections that are small in magnitude but crucial in estimating the invisible walls in the L-NLOS rooms. This indirectly indicates that RGI-Net utilizes higher-order reflections when some low-order reflections are missing.

%% FIG grad-cam
\begin{figure}[t]
\begin{center}
\includegraphics[width=\columnwidth]{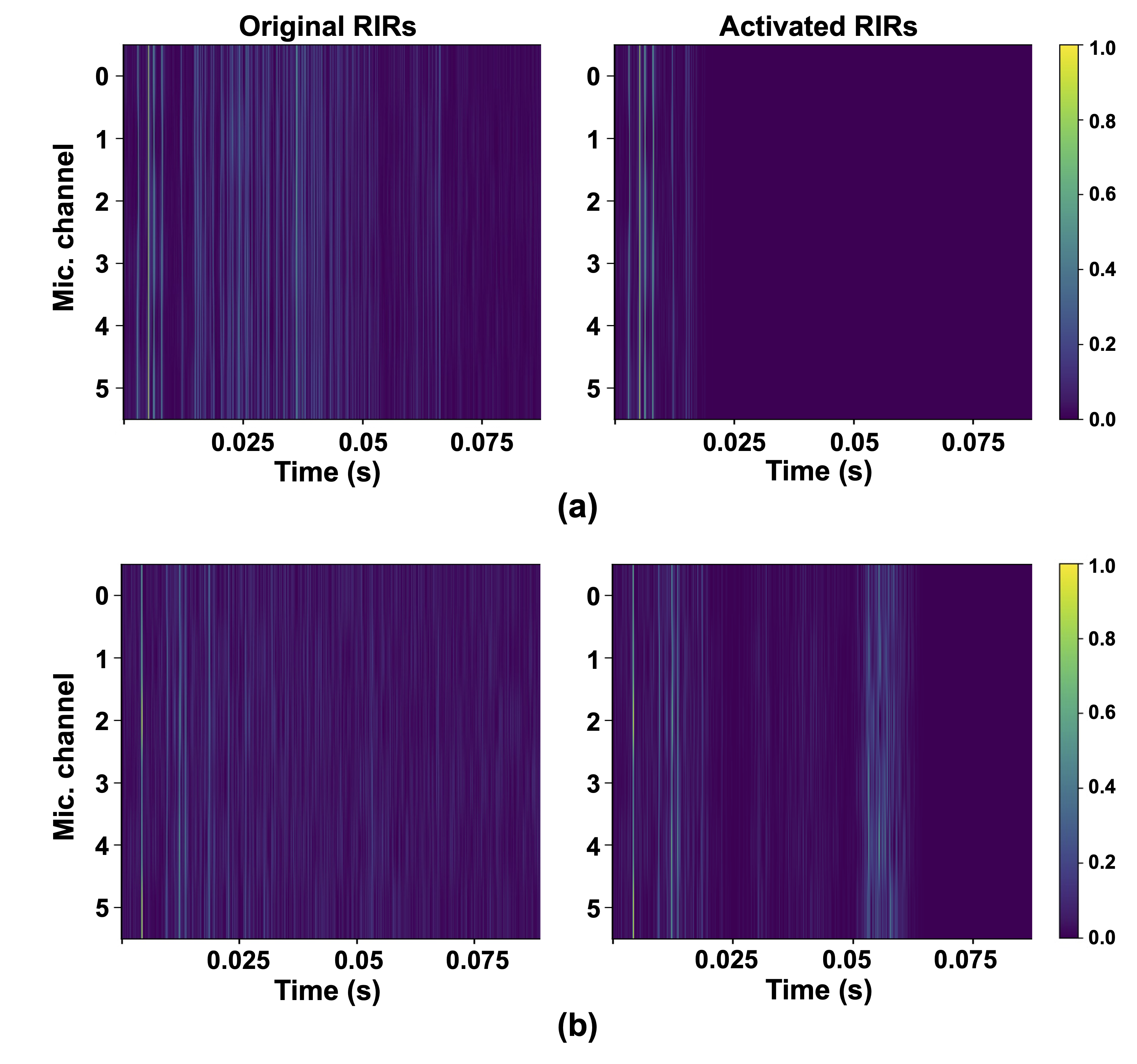}
\caption{Activation maps of multichannel RIRs displaying the use of high-order reflections for geometry inference. (a) convex pentagonal room and (b) non-convex L-NLOS room.}
\label{fig:gradcam}
\end{center}
\vspace{-0.7cm}
\end{figure}

To further verify that RGI-Net can exploit high-order reflections, we generate an activation map using gradient information flowing into the last convolutional layer \cite{gradcam}. In the simple convex pentagonal room (Fig. \ref{fig:gradcam}(a)), the activation map of RGI-Net emphasizes early reflections within the traveling distance of 7 m (20 ms), indicating that low-order reflections are dominantly utilized. In the non-convex L-NLOS room (Fig. \ref{fig:gradcam}(b)), in contrast, the high-order reflections up to 21 m (60 ms) have strong activation. These results visually demonstrate that RGI-Net actively exploits high-order reflections to secure reliability when low-order reflections cannot be seen from the device due to occlusion by other walls.

Next, we describe the RGI performance reported from two conventional \cite{25_baba,28_lovedee} and one DNN-based \cite{tuna_datadriven} methods. The errors of the proposed method shown in Table~\ref{tab:result_compare} are in a similar range to those of the conventional and DNN-based techniques under low-noise conditions. However, this table is not for direct comparison across the methods since there are differences in their room setup, source-microphone configurations, and assumptions. More importantly, the inference of non-convex rooms without relocation of the audio device to secure the LOS condition, and without prior knowledge of the number of walls is a key ability of RGI-Net.

To check the generalizability of the model, we tested the model using RIRs simulated by a different modeling technique. Unlike ISM used for the training data, we simulated test data using ray-tracing with a scattering coefficient of 0.1. The test was conducted without fine-tuning, and three different models trained on clean, low-noise, and high-noise ISM datasets were tested and compared. 
Table \ref{tab:domain_shifted} summarizes the performance variation of three models against the ray-tracing dataset. The model trained on the high-noise dataset exhibits the smallest difference, whereas the model trained with clean RIRs exhibits significantly reduced performances in all metrics. Accordingly, exposure to noise during training helps secure robustness against the different simulation methods.

\begin{table}[t]
\caption{Performance comparison to different RGI methods.}
\centering
\resizebox{\columnwidth}{!}
{
\huge
\begin{tabular}{|l|cccc|cc|cc|}
\hline
\multicolumn{1}{|c|}{\multirow{2}{*}{\backslashbox{Metric}{Method}}} & \multicolumn{4}{c|}{Shoebox} & \multicolumn{2}{c|}{Convex} & \multicolumn{2}{c|}{Non-convex} \\
\multicolumn{1}{|c|}{}   & Ours       & \cite{25_baba}    & \cite{28_lovedee}             & \multicolumn{1}{l|}{\cite{tuna_datadriven}} & Ours       & \cite{28_lovedee}             & Ours       & \cite{28_lovedee}    \\ \hline
$\Delta d$ (m)           & 0.07          & 0.08 & 0.05 & 0.10                    & 0.08          & 0.04 & 0.13 & 0.15 \\
$\Delta \theta$ (degree) & 1.68 & 2.49 & 8.59          & 2.58                    & 2.08 & 5.30          & 1.61 & 5.59 \\ \hline
\end{tabular}%
}
\label{tab:result_compare}
\vspace{-0.3cm}
\end{table}

%% Table 3
\begin{table}[t]
\caption{Accuracy and robustness of the proposed method against noises.}
\centering
\resizebox{\columnwidth}{!}%
{
\Huge
\begin{tabular}{|l|c|c|c|}
\hline
\backslashbox{Metric}{Model} &
  \multicolumn{1}{c|}{\begin{tabular}[c]{@{}c@{}}Model trained with\\ clean RIRs\end{tabular}} &
  \multicolumn{1}{c|}{\begin{tabular}[c]{@{}c@{}}Model trained with\\ low-noise RIRs\end{tabular}} &
  \multicolumn{1}{c|}{\begin{tabular}[c]{@{}c@{}}Model trained with\\ high-noise RIRs\end{tabular}} \\ \hline
\begin{tabular}[c]{@{}l@{}}$ACC_W$ (\%)\\ (ROC--AUC, \%) \end{tabular} &
  \begin{tabular}[c]{@{}c@{}}99.96 $\rightarrow$ ~95.07\\ (99.99 $\rightarrow$ ~94.91) \end{tabular} &
  \begin{tabular}[c]{@{}c@{}}99.95 $\rightarrow$ ~99.45\\ (99.99 $\rightarrow$ ~99.94) \end{tabular} &
  \begin{tabular}[c]{@{}c@{}}99.72 $\rightarrow$ $~\textbf{99.80}$\\ (99.98 $\rightarrow$ $~\textbf{99.98}$) \end{tabular} \\
$\Delta d$ (m)           & 0.10 $\rightarrow$ ~0.41 & 0.10 $\rightarrow$ ~0.17 & 0.16 $\rightarrow$ $~\textbf{0.17}$ \\
$\Delta \theta$ (degree) & 1.82 $\rightarrow$ ~7.26 & 1.89 $\rightarrow$ ~4.04 & 3.33 $\rightarrow$ $~\textbf{3.47}$ \\ \hline
\end{tabular}%
}
\label{tab:domain_shifted}
\vspace{-0.3cm}
\end{table}

\section{Conclusion}
We proposed RGI-Net estimating geometries of complex and non-convex rooms without prior information about the number of walls. This ability was gained by training the model to simultaneously estimate the wall presence probability and geometric parameters of walls. To this end, wall parameter loss and decision loss were defined and used for the model training, which resulted in sufficiently small distance and angular errors even for the RIRs from L-NLOS rooms contaminated by high-level noises. RGI-Net utilizes high-order reflections when first-order reflections cannot be measured from the audio device, which was demonstrated through the visualization of temporal activation maps on RIRs of L-NLOS rooms.

% References should be produced using the bibtex program from suitable
% BiBTeX files (here: strings, refs, manuals). The IEEEbib.bst bibliography
% style file from IEEE produces unsorted bibliography list.
% -------------------------------------------------------------------------
\vfill\pagebreak
\bibliographystyle{IEEEbib}
\bibliography{references}

\end{document}